\documentclass{article}
\usepackage{latexsym,amssymb,amsfonts,amsmath}
\usepackage{graphicx}
\usepackage{caption}
\usepackage{subcaption}
\usepackage[update,prepend]{epstopdf}
\DeclareGraphicsExtensions{.eps,.pdf,.png,.jpg,}
\usepackage{float}
\usepackage{color}
\usepackage[numbers]{natbib}

\title{Bifurcation analysis of the dynamics of interacting populations of spiking networks}
\author{Fereshteh Lagzi, Fatihcan M.\ Atay, Stefan Rotter}
\date{}
\begin{document}
\maketitle

\section*{Abstract}
We analyze the collective dynamics of hierarchically structured networks of densely connected spiking neurons. These networks of sub-networks may represent interactions between cell assemblies or different nuclei in the brain. The dynamical activity pattern that results from these interactions depends on the strength of synaptic coupling between them. Importantly, the overall dynamics of a brain region in the absence of external input, so called ongoing brain activity, has been attributed to the dynamics of such interactions. In our study, two different network scenarios are considered: a system with one inhibitory and two excitatory subnetworks, and a network representation  with three inhibitory subnetworks. To study the effect of synaptic strength on the global dynamics of the network, two parameters for relative couplings between these subnetworks are considered. For each case, a co-dimension two bifurcation analysis is performed and the results have been compared to large-scale network simulations. Our analysis shows that Generalized Lotka-Volterra (GLV) equations, well-known in predator-prey studies, yield a meaningful population-level description for the collective behavior of spiking neuronal interaction, which have a hierarchical structure. In particular, we observed a striking equivalence between the bifurcation diagrams of spiking neuronal networks and their corresponding GLV equations. This study gives new insight on the behavior of neuronal assemblies, and can potentially suggest new mechanisms for altering the dynamical patterns of spiking networks based on changing the synaptic strength between some groups of neurons.



\section*{Introduction}
Networks of pulse-coupled units that operate with a threshold mechanism abound. Examples are forest fires, swarms of flashing fireflies, earthquakes and interacting spiking neurons. Representing the interactions in such systems by a directed graph, each node on the graph receives inputs from many other nodes. If the system input crosses a threshold, a pulse-like signal is emitted and transmitted to the neighboring nodes. The collective behavior of such systems is of particular interest for various reasons, for example, the ability to control the dynamics, or to predict certain events. For instance, in translational neuroscience, one needs to understand the circumstances under which runaway brain states emerge (like an epileptic seizure), or control pathological activity dynamics (like basal ganglia oscillations in Parkinson's disease).    

It is believed that the neocortex has a modular structure with modules that are similar in overall design and operation but different in cell types and connectivity \cite{Mountcastle1997}. This conceptualizes the brain as a hierarchical network of subnetworks that interact with each other, and as a consequence, build a functional brain. Moreover, due to spike timing synaptic plasticity, pre-synaptic neurons that fire together within a close time frame with post-synaptic neurons strengthen synaptic connections that eventually results in formation of cell assemblies \cite{Hebb1949}. Neurons in these assemblies can be connected via short or long range synapses. The interaction of these assemblies may shape an ongoing brain activity that exists even in the absence of external inputs, and may correlate with some internal cognitive states \cite{Harris2005}. This also entails the formation of associative memory \cite{Lansner2009} or synfire chains \cite{Abeles1991,Abeles1982}. On the simulation level, it has been shown that a combination of plasticity mechanisms can lead to input-dependent formation of cell-assemblies \cite{kumar,Tetzlaff2015,Gallinaro2018} which can play role in nonlinear computations.

In subcortical regions, specifically the basal ganglia are comprised of subnetworks (``nuclei'') whose neurons are often conceived as threshold units, sharing similar properties. The synaptic interaction between these subnetworks results in a host of different behaviors, which can often be correlated with either healthy or pathological state. The basal ganglia are connected to many other parts of the brain, including the neocortex and the thalamus. ``Up-states'' and ``down-states'', as examples of dynamical states of neuronal networks, have previously been reported in striatum \cite{Wilson1972,Stern1998}. Cooperation or competition between the different subnetworks of the basal ganglia and certain cortical regions can be conceived as interacting subnetworks of excitatory and inhibitory neurons \cite{Kumar2011,Hammond2007} or only inhibitory neurons \cite{Angulo-Garcia2016,Ponzi2010}. Diseases such as Parkinson's or Huntington's are related to dysfunctions in one or several of these subnetworks, or the interactions between them. Therefore, understanding the role each subnetwork plays in the dynamics of the large network of interconnected brain regions is important to eventually dissect the pathophysiology of these dysfunctional states. Devising novel therapies to alleviate or entirely abolish such pathologies depends on this insight.

To understand the nature of the global dynamics that emerges from these interactions, a theoretical framework is needed. There are different approaches to study the collective dynamics of networks, depending on the exact system in question. Mean-field methods and linear models are common routes taken in the study of the low-dimensional dynamics of spiking neuronal networks \cite{VanVreeswijk1996a,Brunel2000,Gerstner1995,Schwalger2017}. When nonlinearities are important, Wilson-Cowan equations \cite{Wilson1972} are a prime candidate for such analysis.  

In this manuscript, we suggest an alternative low-dimensional firing rate equation for populations of interacting spiking neurons with block-random connections. Originally, these equations were suggested to describe population dynamics in simple ecosystems, with only few interacting species, known as predator-prey dynamics \cite{Volterra1926,lotka}. The analogy between predator-prey systems and spiking neuronal networks is rooted in the competition for limited resources and survival. Prey increases its own population size by reproduction, and also increases the size of the predator population by feeding them. Predators, on the other hand, decrease the prey's population size by feeding on them. A more subtle and more indirect type of interaction is the competition between different predator species that feed on the same type of prey, which results in decreased population sizes of the predators. Likewise, in spiking network dynamics, high activity of excitatory neurons results in a larger number of neurons that are susceptible of firing.
This, in turn, results in a higher activity level of both excitatory and inhibitory populations. Inhibitory neurons, on the other hand, bring down the membrane potential levels of other neurons in the network, and as a consequence, reduce the probability of spike emission. Depending on the feedback structure, this could then lead to a reduction of global network activity. We will show that Generalized Lotka-Volterra (GLV) equations, can capture such collective dynamics quite well, and represent bifurcation diagrams that are validated by spiking network simulations. In other words, with numerical simulations of spiking networks with different coupling parameters, and different structures, we show that there is a qualitative equivalence between these population equations and the steady state behavior of networks of subnetworks of spiking neurons. 

GLV equations have been suggested before as an abstract model of neuronal dynamics \cite{Fukai1997a,Varona2002,Afraimovich2004,Cardanobile2010}, but a systematic generalization to ``networks of interacting networks'' of spiking neurons still needs to be devised. In \cite{Cardanobile2011}, the special case of solutions to the homogeneous GLV equations has been compared to the steady state behavior of point processes with excitatory and inhibitory couplings. GLV equations are known for their rich repertoire of nonlinear dynamics, such as stable pattern formation \cite{Grossberg}, continuous attractors \cite{Li2014}, oscillations \cite{Hofbauer1994}, and chaos \cite{gilpin, Varona2002}. For the locust olfactory system, for example, it has been convincingly demonstrated that the transient dynamics as produced by GLV models can robustly encode sensory information \cite{Rabinovich2008, rabin2001}. It has been hypothesized that sensory representations are more robustly encoded in the transient dynamics, as compared to the stationary activity of neurons in the olfactory system of locusts \cite{mazor2005}, and this behavior is very well captured by GLV equations.

In our study, we compare the behavior of GLV systems to the behavior of networks of subnetworks of excitatory and inhibitory neurons, while coupling parameters between them are changed (bifurcation parameters). We perform a systematic bifurcation analysis of two different networks, each one composed of three subnetworks. An excitatory sub-population has a positive impact on its own activity (growth rate), and hence the analysis is different compared to most GLV studies so far. In other studies, the self-influence of the prey population is negative to reflect a limit on the amount of available resources. It will be shown that the stability of different fixed points of the GLV system depends on the coupling parameters. This dependence is similar to the stable behavioral changes of the corresponding spiking network in the normalized parameter space.
Moreover, in a network with purely inhibitory interactions that follows the May-Leonard structure, oscillations of subnetworks emerge for some parameter regimes, as predicted by the bifurcation diagram of such a system. 

As a model for the collective dynamics of networks of spiking networks, GLV equations also provide us with useful intuition on how to control the dynamics of such networks, with the goal to reestablish a desired behavior, for example in diseased brains.

\section*{Materials and methods}

We studied networks with leaky-integrate-and-fire neuron models with a membrane time constant of $\tau = 20\,\mathrm{ms}$, and a reset potential of $10\,\mathrm{mV}$. Each neuron receives an additional external direct current input of $270\,\mathrm{pA}$. We decided to use a direct current (instead of the commonly used stationary Poisson spike trains) as an external input to each neuron because, for our study, any possible source of random symmetry breaking was to be avoided. 

Each neuron $i$, in a network composed of $N$ neurons, obeys the equation
\begin{equation}
  \tau \dot{v}_i(t) = -v_i(t) + \tau \sum_{j=1}^{N}J_{ij} S_{j}(t) + R I
  \label{LIFeq}
\end{equation}
where $S_j(t)$ is the spike train of a pre-synaptic neuron $j$, which projects to the post-synaptic neuron $i$. In this equation, $v_i$ is the membrane potential of neuron $i$, and $J_{ij}$ is the amplitude of the post-synaptic potential (PSP) caused by spikes in neuron $j$ impinging on neuron $i$. We considered PSPs of amplitude $0.09\,\mathrm{mV}$ for excitatory synapses. In Eq~\eqref{LIFeq}, $I$ is the external DC drive, and $R$ is the input resistance of the neuron. In our simulations of network activity, to regard causality, a uniform synaptic transmission delay of $t_d = 0.1\,\mathrm{ms}$ was used, coinciding with the step size $dt$ for all network simulations. 

All networks studied in this paper are randomly connected. The parameter $\epsilon$ represents the probability of connection between any two neurons within an excitatory subnetwork. The number $p\epsilon$ is the probability of connection between any two inhibitory neurons, or one excitatory and one inhibitory neuron. We chose $p = 3$ to bring the network close to a cortical column \cite{Perin2011}. 

\subsection*{Network structure}

We studied two different scenarios, each of which was characterized by three interacting subnetworks. First, we considered a network, composed of two excitatory subnetworks and one inhibitory subnetwork (EEI). Second, a network of three interacting inhibitory subnetworks was considered (III), which is a neuronal implementation of the May-Leonard system \cite{mayleonard75}. 

\begin{figure}[tbph]
\centering
 \includegraphics[width=0.9\textwidth]{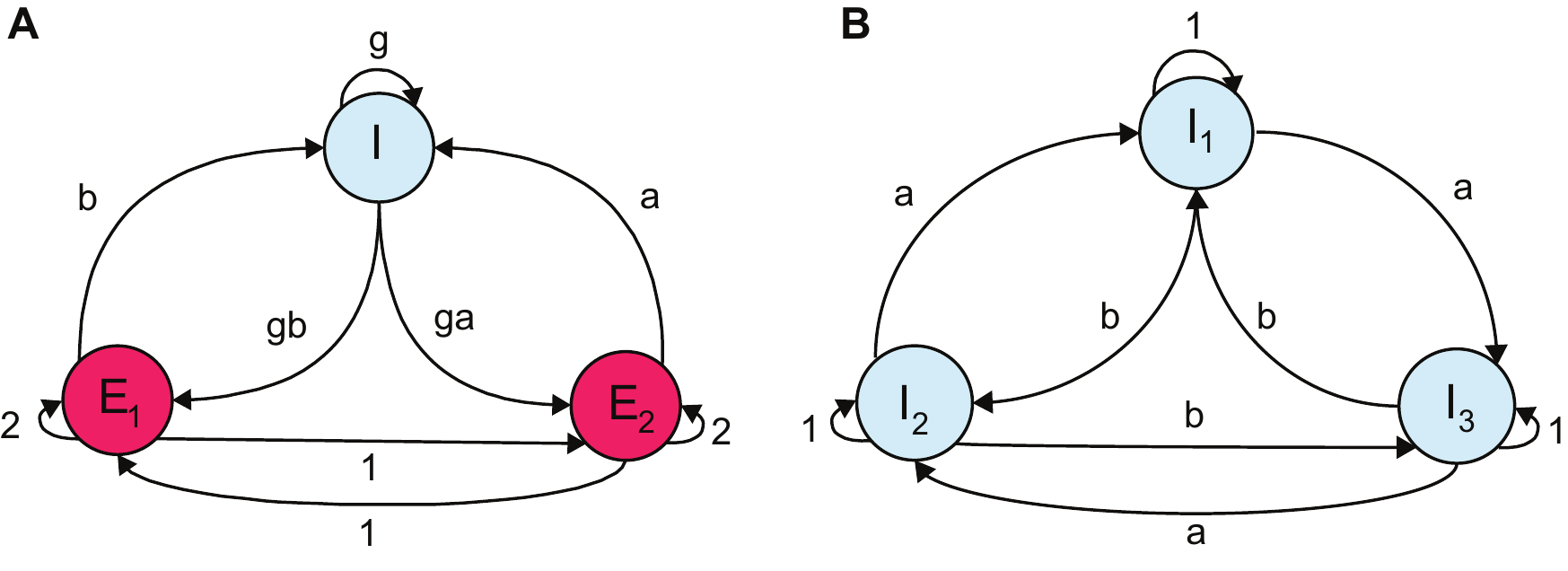}
 \caption{{\bf Two scenarios of a network of subnetworks.} A: EEI scenario, where one inhibitory and two excitatory populations are interacting. B: III scenario, where all the subnetworks are composed of inhibitory neurons. $a$ and $b$ are considered as bifurcation parameters. According to Dale's principle, all connection weights that emanate from inhibitory neurons are negative.}
 \label{fig1}
\end{figure}

For a coarse-grained description, we sum over the dynamics (Eq~\eqref{LIFeq}) of the membrane potentials of all individual neurons in each subnetwork.  
\begin{equation}
\begin{aligned}
  \tau \dot{V}_m(t) &= -V_m(t) + \tau \sum_{n=1}^{3}W_{mn} R_{n}(t) + N_m r_0(t) \\
  &= -V_m + L_m(R).
  \label{SubV}
  \end{aligned}
\end{equation}
In this equation, $V_m$ is the sum of the membrane potentials of the neurons in population $m$. $R_n$ is is the sum of spike trains of neurons in population $n$. In other words, its mathematical expectation is the collective firing rate of population $n$. Variable $r_0(t)$ is the equivalent firing rate for the external DC input to each neuron, and $N_m$ is the size of the subnetwork $m$. $L_m(.)$ is a function representing a linear combination of the firing rates of all subnetworks, as well as the external firing rate to each subnetwork $m$. It is easily verified that for a network with fixed subnetwork-specific out-degrees for each neuron, the connectivity matrix $W$ for the EEI network is
\[
W_\mathrm{EEI} = \tau \epsilon J  
 \begin{pmatrix}
  w N_E & N_E & -p g b N_E\\
  N_E & w N_E & -p g a N_E\\
  p b N_I & p a N_I & -p g N_I
 \end{pmatrix}
\]
where $w$ is a factor describing the relative PSP amplitude for couplings within excitatory populations. Inspired by \cite{Perin2011} where they showed that clustered excitatory neurons build up stronger EPSP amplitude, we chose $w=2$ for the simulations. The parameters $a$ and $b$ are the scaling weight parameters, affecting the strength of neuronal connections. We refer to them as bifurcation parameters. The parameter $g$ is the amplitude ratio between IPSPs (inhibitory post-synaptic potentials) and EPSPs (excitatory post-synaptic potentials). For the III scenario, we considered the following coupling matrix, according to \cite{mayleonard75}:
\[
W_\mathrm{III} = \tau \epsilon g J N_I
 \begin{pmatrix}
  -1 & -a & -b\\
  -b & -1 & -a\\
  -a & -b & -1
 \end{pmatrix}
\]

\subsection*{Exponential transfer function}

To obtain firing rate equations for the activities of subnetworks that are involved in a network, we need to know the dynamic transfer function between the collective membrane potential of each subnetwork and it's corresponding firing rate. In \cite{Wilson1972}, it was hypothesized that the response function of a sub-population to the available amount of excitation in the network, in the simple case of a unimodal distribution of synaptic weights, could be well approximated by a sigmoid function (for a more general case, see \cite{Gerstner1995}). A study on transient dynamics of balanced random networks showed that in low firing rate regimes, the distribution of the collective excitatory and inhibitory spike counts follows a log-normal distribution \cite{Lagzi2014}. Assuming in this regime, the neuronal membrane potentials states could be approximated by a normal distribution \cite{Brunel2000}, this observation implies that an exponential transfer function links the collective membrane potential of each population to its firing rate. Motivated by this idea, here, we assume that if the network is in the low firing rate regime, and the neuronal refractory period can be neglected, the relationship between the collective membrane potential and firing rate of neurons can be approximated by an exponential function 
\begin{equation}
 R = \alpha \exp{(\beta V)}.
 \label{EXP_func}
\end{equation}
In general, the parameters of this function depend on the size of the neuronal population, as well as on the parameters of the neurons. However, for the sake of simplicity, we assume that they are identical for the three subnetworks in this study.

\subsection*{Lotka-Volterra equations}

Equation~\eqref{SubV} together with the exponential relationship between $V_s$ and $R_s$ given by Eq~\eqref{EXP_func} for a subnetwork $s$ yields the following relationships
\begin{equation}
\begin{aligned}
 \dot{R_s} &= \alpha \beta \dot{V_s} \exp{(\beta V_s)} = \beta R_s \dot{V_s} \\
 &= R_s(-\frac{1}{\tau}\log{\frac{R_s}{\alpha}} + \frac{\beta}{\tau} L_s(R)).
 \label{LVeq}
 \end{aligned}
\end{equation}
As the coefficients of the linear term are much larger than $1$, the logarithmic term in Eq~\eqref{LVeq} makes a negligible contribution and will be omitted in our analysis. The resulting equation is a Generalized Lotka-Volterra (GLV) equation. The parameters of the linear function $L_a(R)$ are inferred from the corresponding connectivity matrix, $W_\mathrm{EEI}$ or $W_\mathrm{III}$. It is important to note that Lotka-Volterra equations represent collective dynamics, regardless of the dynamics of individual nodes. Therefore, the procedure in this paper is to confirm the applicability of such equations, particularly to reduce the dimensionality of the dynamics in networks of leaky integrate-and-fire (LIF) neurons. 

\subsubsection*{EEI scenario}

In this case, two excitatory subnetworks of size $N_E=6\,000$ each are reciprocally connected to an inhibitory population of size $N_I=3\,000$ (Figure~\ref{fig1}A). The network is constructed in such a way that for each subnetwork, neurons have identical in-degrees and the number of outgoing connections to each subnetwork are the same (configuration model \cite{Newman2001,Hofstad2014}). The rate equations for the involved populations follow
\begin{equation}
\begin{aligned} 
   \dot{x}_1(t) &= k x_1(t)(2 w x_1(t) + 2 x_2(t) -2 p g b y(t) + 2 I)  \\
   \dot{x}_2(t) &= k x_2(t)(2 x_1(t) + 2 w x_2(t) -2 p g a y(t) + 2 I ) \\
   \dot{y}(t) &= k y(t)(b p x_1(t) + a p x_2(t) -p g y(t) + I ) 
\end{aligned}
 \label{LV_EEI}
\end{equation}
where $x_1$ and $x_2$ are the firing rates of the excitatory populations and $y$ stands for the firing rate of the inhibitory population. As mentioned before, it is assumed that the couplings within each excitatory population are twice stronger than couplings between excitatory neurons in different subnetworks ($w=2$). The factor $2$ inside the linear part of the first two equations in~\eqref{LV_EEI} reflects the fact that $N_E = 2 N_I$ (the reader is referred to $W_\mathrm{EEI}$ and $W_\mathrm{III}$ in the ``Network structure'' section). The parameter $I$ represents the external input to each population. For simplicity, we consider $I = 1$. This reflects the fact that the external input current was close to the rheobase of the neurons in the network ($250\,\mathrm{pA}$). Parameter $k=1$, without loss of generality, represents the time-scale of the system (a new time variable can be defined by rescaling $t$). 

To study the time-dependent dynamics and the steady state behavior of the overall network, it is important to analyze the fixed point solutions of Eq~\eqref{LV_EEI} and their stability properties. A three-dimensional GLV, like Eq~\eqref{LV_EEI}, typically has $2^3 = 8$ fixed points, corresponding to zero or non-zero solutions of the three dynamical variables $x_1$, $x_2$, and $y$. Here, symbolically, we denote the zero and non-zero solutions by $0$ and $1$, respectively, although non-zero solutions are not necessarily numerically equal to $1$. We show the solutions and their corresponding system eigenvalues as a function of the bifurcation parameters $a$ and $b$. In the following, we use the parameters $g = 6$, $p = 3$, and $w = 2$.

For the fixed point corresponding to $(x_1^*,x_2^*,y^*)=p_{1,1,1}$, where $p_{1,1,1}$ indicates the fixed point with all nonzero solutions of equation \eqref{LV_EEI}, the parametric solutions are
\[
\begin{bmatrix}
  x_1^*\\[0.3em]
  x_2^*\\[0.3em]
  y^*
 \end{bmatrix} = \frac{-I}{3(-2a^2+2ab-2b^2+1)}
 \begin{bmatrix}
  a-2b+3ab-3a^2+1\\[0.3em]
  b-2a+3ab-3b^2+1\\[0.3em]
  \frac{a+b-1}{6}
 \end{bmatrix}.
\]
In order to study the local stability of this fixed point, the eigenvalues of Jacobian matrix evaluated at the fixed point need to be considered. The Jacobian matrix for $p_{1,1,1}$ is
\[
Jac_\mathrm{111} =
 \begin{bmatrix}
  8x_1^* + 2x_2^* -36by^*+2I & 2x_1^* & -36bx_1^*\\
  2x_2^* & 2x_1^*+8x_2^*-36ay^*+2I & -36ax_2^*\\
  3by^* & 3ay^* & 3bx_1^*+3ax_2^*-36y^*+I
 \end{bmatrix}.
\]
Numerical continuation method of integration \cite{Dhooge2003} for these equations show that for $p_{1,1,1}$, there is a supercritical Hopf bifurcation that passes through $p_{0,0,1}$. This bifurcation line is illustrated in red in Figure~\ref{fig2}A.

For the solution $(x_1^*,x_2^*,y^*)=p_{0,1,1}$, it is easy to get
\[
\begin{bmatrix}
  x_1^*\\[0.3em]
  x_2^*\\[0.3em]
  y^*
 \end{bmatrix} = 
 \begin{bmatrix}
  0\\[0.3em]
  \frac{I(1-a)}{3a^2-2}\\[0.3em]
  \frac{I(3a-2)}{18(3a^2-2)}
 \end{bmatrix}.
\]
The eigenvalues of the Jacobian at this fixed point are
\begin{equation}
\begin{aligned} 
   \lambda_1 &= \frac{2I(a-2b+3ab-3a^2+1)}{2-3a^2}  \\
   \lambda_{2,3} &= \frac{I(6-7a \pm \sqrt{72a^4 -120a^3 +49a^2 -4a +4})}{6a^2-4} 
\end{aligned}
\end{equation}
To obtain the transcritical bifurcation lines \cite{strogatz2000} in the parameter space, one needs to solve for $\lambda_1 = 0$ and $\lambda_{2,3} = 0$. The former results in $b = \frac{3a^2-a-1}{3a-2}$, which is a curve in the $a-b$ plane. The latter will result in $a = 1$ and $a = 2/3$ for the transcritical bifurcation lines, and $a = 0.8571$ for a degenerate Hopf bifurcation. Due to the symmetry between $x_1$ and $x_2$ in the equations, it is easy to get the solutions for the fixed point $(x_1^*,x_2^*,y^*)=p_{1,0,1}$. In this case, $a = \frac{3b^2-b-1}{3b-2}$, as well as $b = 1$, $b = 2/3$, and $b = 0.816$ determine a transcritical bifurcation line. The line $b = 0.8571$ represents a degenerate Hopf bifurcation for this fixed point.

For $(x_1^*,x_2^*,y^*)=(0,0,1)$, it turns out that the values of the parameters do not play any role in determining the fixed point which is $(0,0,I/18)$. The corresponding eigenvalues of the Jacobian are
\begin{equation}
\begin{aligned} 
   \lambda_1 &= -I \\
   \lambda_{2} &= -2I(a-1)\\
   \lambda_{3} &= -2I(b-1)
\end{aligned}.
\end{equation}
This indicates that for $a,b > 1$, the fixed point is locally stable. 

For the origin $(x_1^*,x_2^*,y^*)=p_{0,0,0}=(0,0,0)$, the eigenvalues are $\lambda_1 = I$, and $\lambda_{2,3}=2I$, which are always positive in spiking neural networks with a positive external input. Furthermore, for $(x_1^*,x_2^*,y^*)=p_{1,1,0}$, the parametric solutions are $(-I/3, -I/3, 0)$. For $(x_1^*,x_2^*,y^*)=p_{1,0,0}$ and $p_{0,1,0}$, the parametric solutions are $(-I/2,0,0)$ and $(0,-I/2,0)$, respectively. The last three cases are impossible solutions for the firing rates of spiking neuronal networks. Therefore, we will not consider their stability and their influence on the trajectory of the network.

\subsubsection*{III scenario}

This system represents a neuronal implementation of the May-Leonard equation \cite{mayleonard75}, which is well-known for generating oscillatory population dynamics for some parameter ranges. As depicted in Figure~\ref{fig1}B, the scaling factor for the coupling weights is $a$ for clockwise connections, and $b$ for counterclockwise couplings. In this case, each inhibitory subnetwork comprises of $4\,000$ neurons. The corresponding equations for the GLV dynamics are
\begin{equation}
\begin{aligned} 
   \dot{x}_1(t) &= k x_1(t)(-x_1(t) -a x_2(t) -b x_3(t) + I)  \\
   \dot{x}_2(t) &= k x_2(t)(-b x_1(t) - x_2(t) -a x_3(t) + I ) \\
   \dot{x}_3(t) &= k x_3(t)(-a x_1(t) -b x_2(t) -x_3(t) + I ) 
\end{aligned}
 \label{LV_III}
\end{equation}
where $x_1$, $x_2$ and $x_3$ represent the firing rates of the three inhibitory subnetworks, respectively. Similar to the EEI case, we assume $I=1$. A full bifurcation analysis for this system of equations is given in \cite{mayleonard75}.

\section*{Results}

In this section, we compare our analytical treatment of the GLV equations with spiking network simulations for both examples considered in this paper, i.e.\ EEI and III networks. All numerical simulations were conducted in NEST \cite{Gewaltig2007} for a duration of $4$ seconds.

\subsection*{EEI scenario}

For a generic three-dimensional Lotka-Volterra equation, $2^3=8$ different fixed points are possible. However, in the Lotka-Volterra system with $W_\mathrm{EEI}$ connectivity matrix, three of the fixed points have negative components regardless of the choice of parameters $a$ and $b$, excluding them as a biological firing rate. Consequently, with initial conditions chosen in the positive octant, only the fixed points in this octant need to be considered. The reason is that in a GLV system, trajectories remain non-negative if the initial conditions have this property. Moreover, in Eq~\eqref{LV_EEI} the origin always represents an unstable fixed point. Therefore, we only analyze the system dynamics for four different fixed points $(x_1,x_2,y)$, denoted by $p_{1,1,1}$, $p_{1,0,1}$, $p_{0,1,1}$, $p_{0,0,1}$. Here, $0$ means no activity and $1$, symbolically, means that the corresponding population is active and has a non-zero firing rate. 

\subsubsection*{Stability analysis}

\begin{figure}[tbph]
        \centering
        \includegraphics[width=\textwidth]{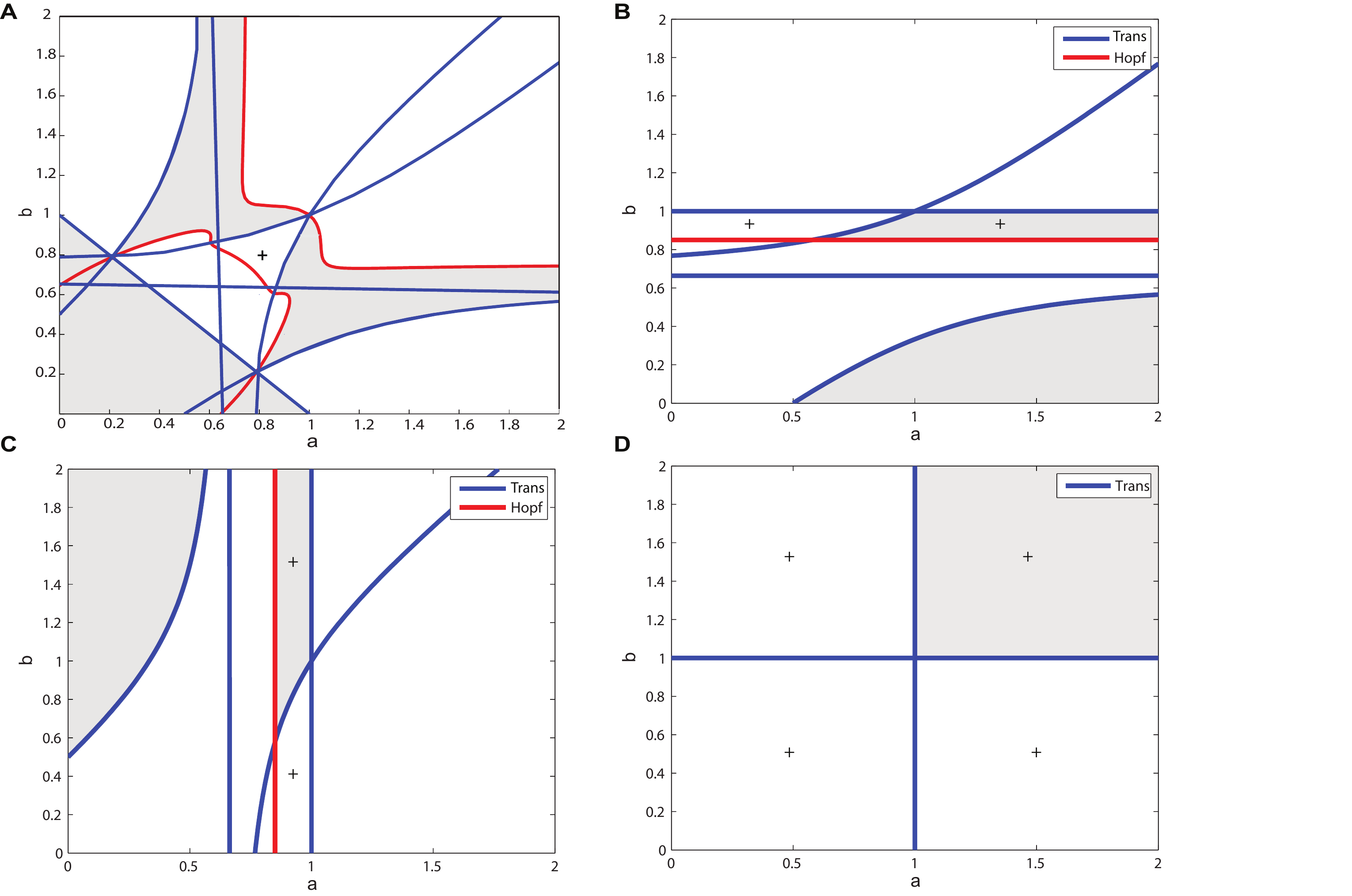}
        \caption{{\bf Local stability analysis of fixed points in the EEI scenario.} Stable regions for A: $p_{1,1,1}$, B: $p_{1,0,1}$, C: $p_{0,1,1}$ and D: $p_{0,0,1}$ are depicted in gray. The corresponding fixed point in the white regions has at least one unstable manifold (at least one eigenvalue has a positive real part). In each case, Hopf and transcritical bifurcations are depicted in red and blue, respectively. Regions with a $(+)$ show that the fixed point is in the positive octant of the state space.}
\label{fig2}
\end{figure}

Figure~\ref{fig2} depicts the locally stable regions in the parameter space, where all eigenvalues have a negative real part (gray shaded area), for each of the above-mentioned fixed points. Moreover, Hopf and transcritical bifurcation lines for each fixed point are represented in red and blue, respectively (The numerical bifurcation analysis was performed in MATCONT v6.1 \cite{Dhooge2003}). A $+$ sign in a region indicates that the corresponding fixed point is in the first octant.

The straight lines $a = 0.8571$ and $b = 0.8571$ are generalized Hopf bifurcation for $p_{0,1,1}$ and $p_{1,0,1}$, respectively. This bifurcation is not generic, however, since the second Lyapunov exponent is zero. Numerical analysis shows that for these parameter values, in the first octant and on the $x_1=0$ plane or $x_2=0$ plane, there is one saddle limit cycle, as there are two Floquet multipliers (eigenvalues of the discrete map for the cycle) equal to $-1$ and $+1$. As the value of the parameter increases, the trajectory will converge to a stable fixed point on the plane; meaning, depending on the initial condition, the trajectory converges either to $p_{1,0,1}$ or to $p_{0,1,1}$. 

\begin{figure}[tbph]
      \centering
     \includegraphics[width=0.7\textwidth]{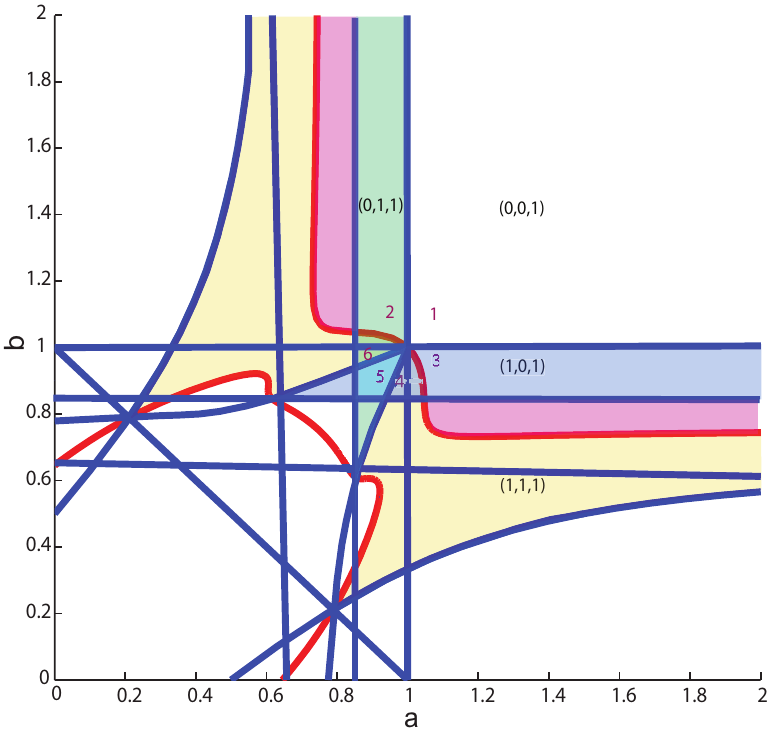}
      \caption{{\bf Local stability of fixed points in the parameter space.} For the region of interest, the stability of different fixed points is highlighted with different colors. See the text for more details.}
\label{fig3}      
\end{figure}

\begin{figure}[tbph]
\centering
     \includegraphics[width=\textwidth]{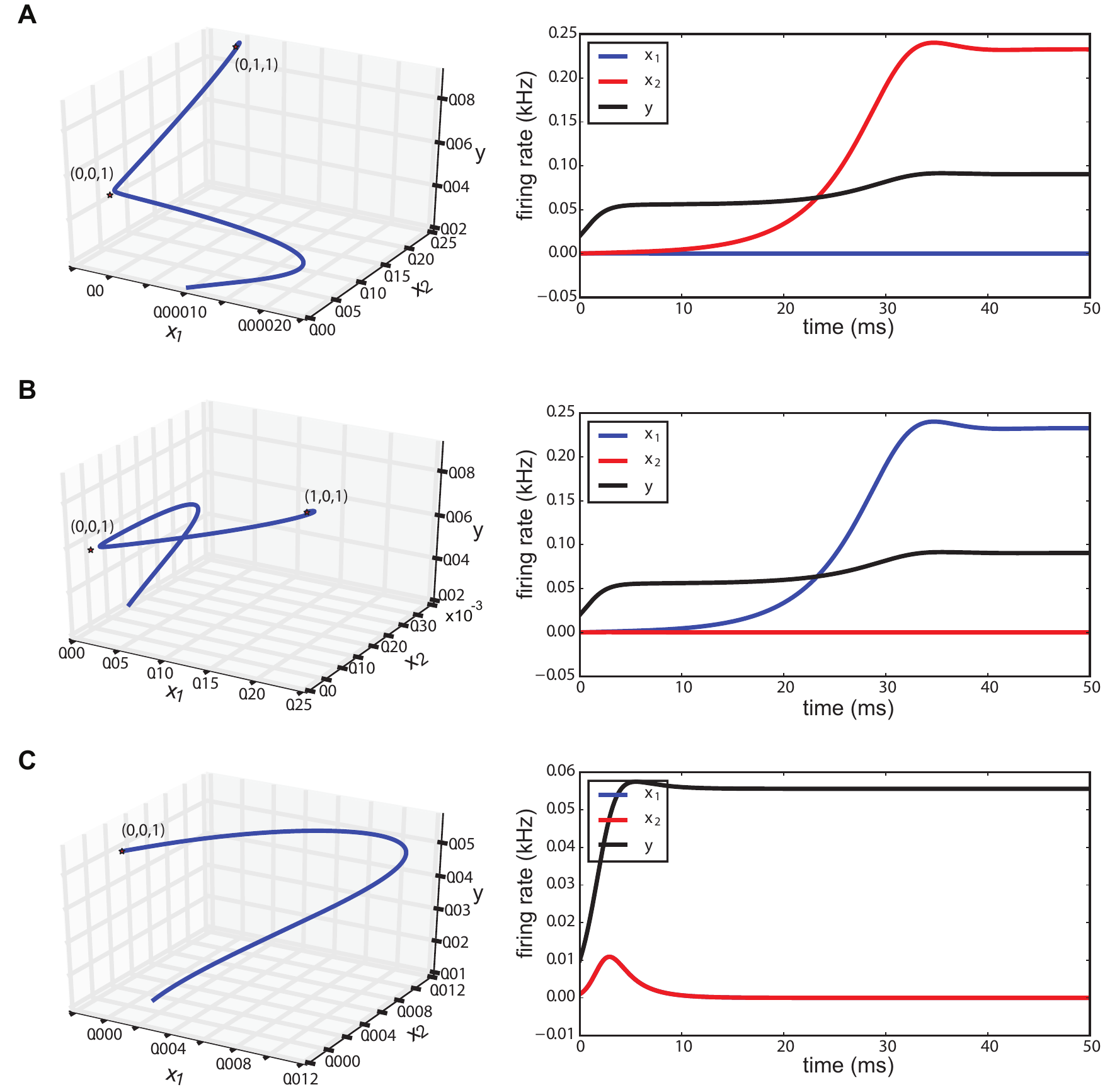}
      \caption{{\bf Numerical integration of Eq~\eqref{LV_EEI} for different parameters corresponding to the $6$ regions indicated in Figure~\ref{fig3}.} A: For $a=0.9$, $b=1.3$ (region $2$) and initial conditions $(0.0001,0.0001,0.02)$, the trajectory passes by the vicinity of $p_{0,0,1}$, and eventually converges to $p_{0,1,1}$. B: For $a=1.2$, $b=0.9$, and $(0.0001,0.0001,0.02)$ as the initial condition, the trajectory follows a heteroclinic connection between $p_{0,0,1}$ and $p_{1,0,1}$. This parameter combination corresponds to region $3$. C: For $a=b=1.2$, and any initial condition in the first octant, the trajectory converges to $p_{0,0,1}$.}
\label{fig4}      
\end{figure}

We now focus on the parameter region $a, b > 0.8571$, because only under this condition neurons operate in the fluctuation-driven regime, indicated by network simulations. Different bifurcation lines of the fixed points divide the parameter space into six different regions (Figure~\ref{fig3}). In region 1, $p_{0,0,1}$ is a globally stable fixed point, and therefore, all trajectories will end in this point. Figure~\ref{fig4}C illustrates a trajectory in the state space (left) and time domain (right) in this parameter space.

In regions 2 and 3 in Figure~\ref{fig3}, apart from the origin $p_{0,0,0}$ and $p_{0,0,1}$, there exist also $p_{0,1,1}$ and $p_{1,0,1}$, respectively. In these regions, there is a heteroclinic connection between $p_{0,0,0}$ to $p_{0,0,1}$. The only unstable manifold of $p_{0,0,1}$ is directed to the stable manifold of $p_{0,1,1}$ in region 2 (Figure~\ref{fig4}A), and to the stable manifold of $p_{1,0,1}$ in region 3 (Figure~\ref{fig4}B). Thus, depending on the initial conditions, the trajectory passes by a few number of saddle points and eventually settles in a globally stable fixed point.  

\begin{figure}[tbph]
\centering
      \includegraphics[width=\textwidth]{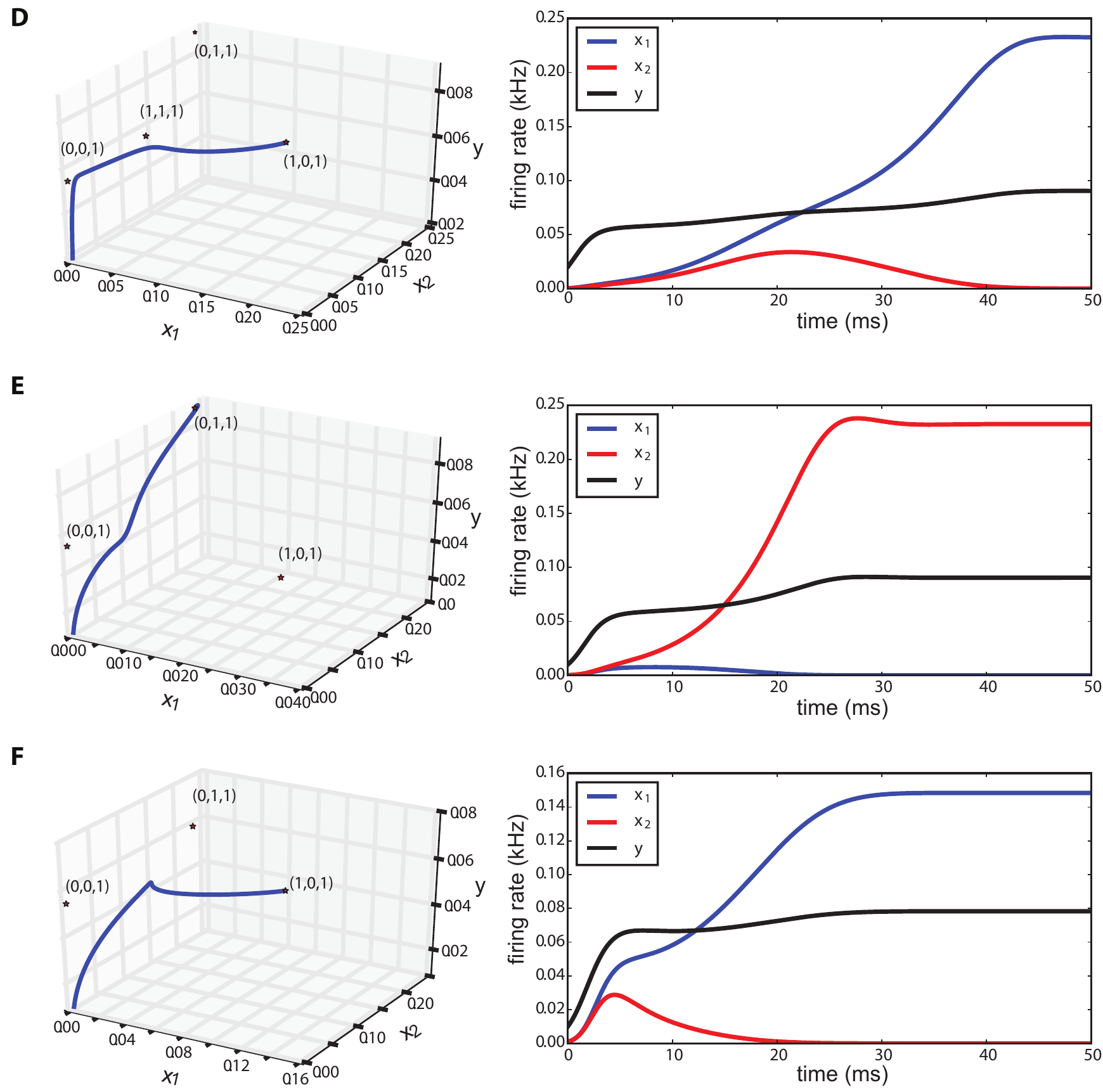}
      \caption{{\bf Numerical integration of Eq~\eqref{LV_EEI} for different parameters corresponding to the $6$ regions.} D: For $a=b=0.9$, (region $5$) and initial conditions equal to $(0.0004,0.0003,0.02)$, the trajectory passes by $p_{0,0,1}$ and $p_{1,1,1}$, and will converge to $p_{1,0,1}$. In this case, a longer heteroclinic connection between saddle points exists. If the initial condition is such that $x_2 > x_1$, the trajectory will converge to $p_{0,1,1}$. E: For $a=0.90$, $b=0.97$, and $(0.0002,0.0002,0.01)$ as the initial condition, the trajectory represents a heteroclinic connection between $p_{0,0,1}$ and $p_{1,0,1}$, and finally will converge to $p_{0,1,1}$. The amount of time that the trajectory spends close to the saddle points, and also its distance to each saddle point, depends on the initial conditions. This parameter combination corresponds to region $6$. F: For $a=0.98$, $b=0.92$ which corresponds to region$4$, and the initial condition $(0.001,0.001,0.01)$, the trajectory follows a heteroclinic connection between $p_{0,0,0}$, $p_{0,0,1}$ and $p_{0,1,1}$, and eventually converges to $p_{1,0,1}$.}
\label{fig5}      
\end{figure}

\begin{figure}[tbph]
        \centering
      \includegraphics[width=\textwidth]{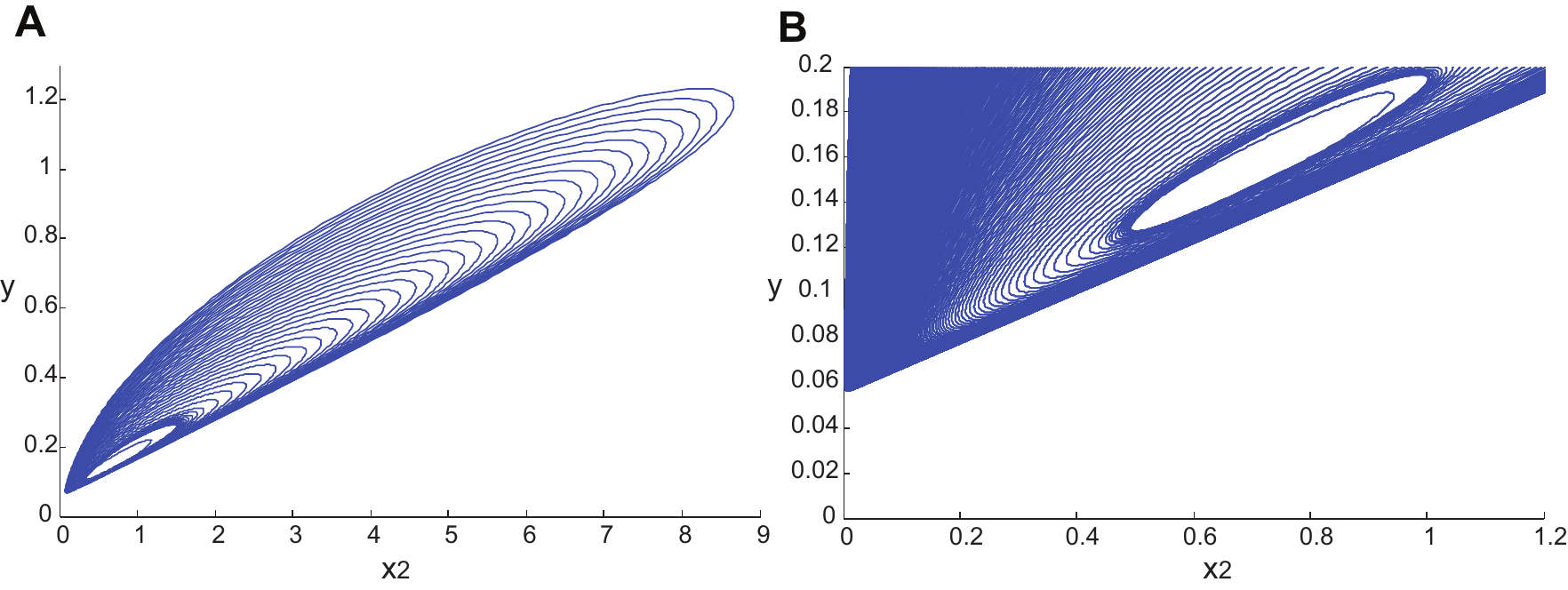}
      \caption{{\bf Homoclinic connections for $p_{0,0,1}$ on the $x_1=0$ plane.} A: Limit cycles that bifurcate from $p_{0,1,1}$ at $a=0.8571$ become tangent to $p_{0,0,1}$ when the period of the cycle tends to infinity. B: Zoom into the area of interest, close to the point $(0,0,0.0556)$ which is the fixed point of the GLV corresponding to $p_{0,0,1}$.}
      \label{fig6}
\end{figure}

In regions 4, 5 and 6, the fixed point $p_{0,0,1}$ has a two-dimensional unstable and a one-dimensional stable manifold with real eigenvalues. The directions of the eigenvectors corresponding to the unstable eigenvalues point towards the other fixed points $p_{0,1,1}$ and $p_{1,0,1}$ and build a heteroclinic connection (Figure~\ref{fig5}). The $y$ axis is the stable manifold. The saddle value, i.e., the sum of the real parts of the closest eigenvalues to the imaginary axis on the right and left side of the axis, for this fixed point is always negative, in the range of interest for $a$ and $b$. 
As pointed out earlier, $a = 0.8571$ and $b = 0.8571$ are generalized Hopf bifurcation curves for $p_{0,1,1}$ and $p_{1,0,1}$, respectively. From our numerical analysis it turns out that the limit cycles that bifurcate from these fixed points touch the $p_{0,0,1}$ fixed point when the period of the cycle tends to infinity (Figure~\ref{fig6}). Therefore, $a = 0.8571$ and $b = 0.8571$ are homoclinic bifurcation curves for $p_{0,0,1}$ as well. In this parameter range, the saddle value is negative. According to Shilnikov's theorem for two unstable--one stable manifold in a three-dimensional system (see \cite{Kuznetsov2004}), the limit cycles that bifurcate from this curve should, under generic conditions, generate a saddle limit cycle in the vicinity of $x_1 = 0$ and $x_2 = 0$ planes when $a$ and $b$ increase. However, the Hopf bifurcation is degenerate and only on the bifurcation lines $a = 0.8571$ and $b = 0.8571$, a homoclinic connection exists. Therefore, as the values of the parameters are changed away from the bifurcation lines, any trajectory will end up in the corresponding fixed point.  

In region 5, depending on the initial conditions, the trajectory will either converge to $p_{0,1,1}$ or $p_{1,0,1}$. For $a=0.9$ and $b=0.9$, for example, both fixed points have two complex conjugate eigenvalues with negative real parts, and one negative real eigenvalue.
Moreover, in this region, $(1,1,1)$ is in the first octant and has one positive real eigenvalue for which the corresponding unstable manifold points towards the fixed points $(1,0,1)$ or $(0,1,1)$ (Figure~\ref{fig5}D). Depending on possible fluctuations in network activity, the trajectory would eventually converge to either of these two fixed points.

In region 6, for $a=0.9$ and $b=0.97$ for example, $p_{0,1,1}$ has three eigenvalues with negative real parts, two of which are complex conjugate. This indicates the local stability of this point. The point $p_{1,0,1}$ has three real eigenvalues, two negative and one positive eigenvalue. In fact, the latter has changed sign due to a bifurcation, as the value of $b$ was increased. In this region, $p_{0,0,1}$ still has two positive real eigenvalues and one negative real eigenvalue, and $p_{1,1,1}$ is not in the first octant. The fixed point $p_{1,0,1}$ has a positive eigenvalue which creates an unstable manifold in the direction of the only attractor in the first octant, namely $p_{0,1,1}$. Putting all this information together, one concludes that there is a heteroclinic chain for the trajectory, starting in $p_{0,0,0}$ and then moving to $p_{0,0,1}$, and finally toward $p_{0,1,1}$ (Figure~\ref{fig5}E). For initial conditions close to $p_{1,0,1}$, the trajectory will be lead to $p_{0,1,1}$ via the unstable manifold of $p_{1,0,1}$. Due to the existing symmetry between $a$ and $b$ in the parameter space, the same behavior applies to region 4, but with the relevant fixed points interchanged (Figure~\ref{fig5}F).

\subsubsection*{Network simulation}

\begin{figure}[tbph]
        \centering
        \includegraphics[width=\textwidth]{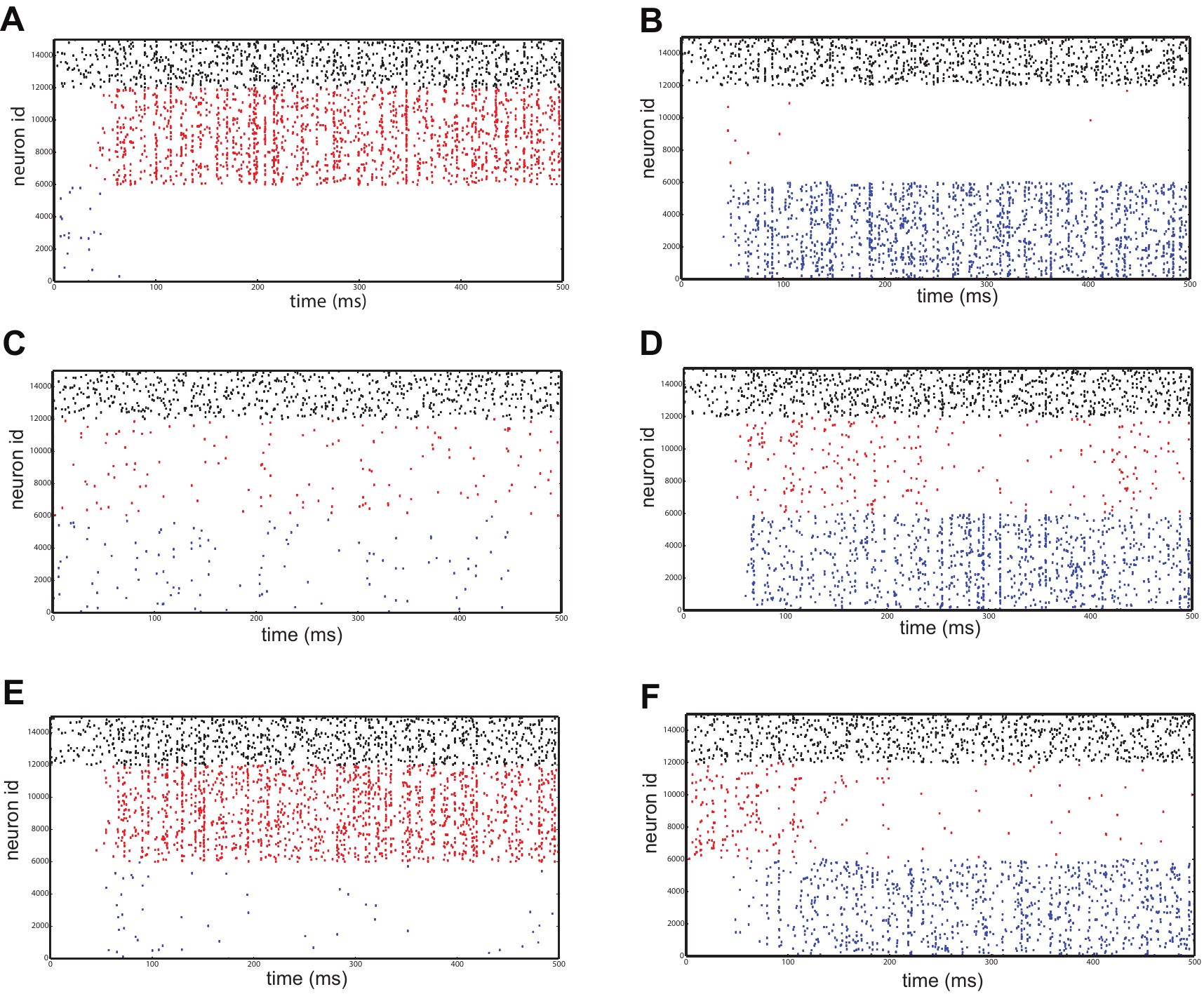}
        \caption{{\bf Raster plots of the network activity for the EEI scenario.} A: For $a=0.9$ and $b=1.3$, the second excitatory population dominates the activity of the first excitatory population. B: For $a=1.2$ and $b=0.9$, the first excitatory population is more active than the second excitatory population. C: For $a=1.2$ and $b=1.2$, the only strongly active population is the inhibitory subnetwork. D: For $a=0.9$ and $b=0.9$, the two excitatory populations compete with each other. It depends on the initial fluctuations of the network dynamics which population wins the competition. E: At $a=0.90$ and $b=0.97$, the network activity is such that the inhibitory population is active together with the second excitatory population. F: for $a=0.98$ and $b=0.92$, starting with an initial condition that favors the second excitatory population, after a short transient, the network activity converges to a state in which the first excitatory population as well as the inhibitory population have higher activity.} 
\label{fig7}
\end{figure}

We performed numerical simulations of networks of leaky integrate-and-fire neurons for samples of parameter combinations $(a,b)$ from the six regions defined in the previous section. Raster plots of the two populations of excitatory neurons are shown in blue and red corresponding to $x_1$ and $x_2$, respectively, in Figure~\ref{fig7}. The activity of neurons from the inhibitory population are shown in black. The index of the neurons on the vertical axis are between $1$ and $15\,000$, the first $6\,000$ neurons belong to the first excitatory population. Neurons with an index between $6\,001$ and $12\,000$ belong to the second excitatory population, and neurons with indices between $12\,001$ and $15\,000$ correspond to the inhibitory population. In all of the sub-figures, after a short transient, the network activity remains in a steady state, which is stable apart from quasi-stochastic fluctuations. 

In Figure~\ref{fig7} A, B, D the initial conditions for the neuronal membrane potentials were randomly chosen between the threshold value at $0\,\mathrm{mV}$ and $15\,\mathrm{mV}$, for the excitatory neurons. The initial conditions for neurons in the inhibitory subnetwork were chosen between $0$ and $20\,\mathrm{mV}$ (the threshold level is at $20\,\mathrm{mV}$ for all neurons). In a reduced-dimensional model, this would correspond to an initial condition close to $p_{0,0,1}$. For details on the initial conditions and the model behavior, see Figure~\ref{fig4} and Figure~\ref{fig5}. In panel~A, in the first $50\,\mathrm{ms}$ of the network simulation, there is some weak activity of the first excitatory population, together with the inhibitory population. Afterwards, the steady state activity is such that the inhibitory population together with the second excitatory population remain active, and the first excitatory population becomes silent. This nicely matches the behavior of the trajectory in region 2, displayed in Figure~\ref{fig4}A. In panel B, the activity in the transient phase corresponds to an initial condition close to $p_{0,0,1}$ (as explained before). After a short time, the inhibitory and first excitatory subnetwork are highly active, and the second excitatory subnetwork remains silent. This was also expected from the GLV analysis (Figure~\ref{fig4}B). In panel D, $a$ and $b$ are both set to $0.9$. The GLV analysis indicates that the saddle fixed point $p_{1,1,1}$ is localized in the first octant. The network activity is such that after a high activity of the inhibitory population and the extended silence period of the two excitatory populations for almost $50\,\mathrm{ms}$, the network activity switches to a state wherein all three populations are active, corresponding to the saddle point $p_{1,1,1}$. After about $150\,\mathrm{ms}$, the network activity settles in a steady state, where the inhibitory subnetwork together with the first excitatory subnetwork are active. This corresponds to the stable fixed point $p_{1,0,1}$ in the GLV equations (Figure~\ref{fig5}D). In repeated network simulations, we observed that the steady state could also correspond to $p_{0,1,1}$, with equal likelihood. Also, for parameters $a$ and $b$ close to $1$, the network can exhibit bistable dynamics (see \cite{lagzi2015} for more details). 

For $a = b = 1.2$, the initial membrane potentials of all neurons, regardless of their synaptic connections, were chosen randomly between rest and threshold ($0$ and $20\,\mathrm{mV}$). This was to investigate the network preference for unbiased initial conditions. It was observed that the inhibitory neuronal population exhibits a considerably higher firing rate as compared to the excitatory subnetworks (Figure~\ref{fig7} C). GLV integration results also illustrate a unique stable fixed point in the first octant, that is $p_{0,0,1}$ (Figure~\ref{fig4}C).

As an example corresponding to region 6 in Figure~\ref{fig3}, we chose $a=0.9$ and $b=0.97$ (Figure~\ref{fig7}E). The initial conditions were chosen such that in the low-dimensional system, the network is close to $p_{0,0,1}$. It was observed that, after some time, the activity of the network reaches a steady state, where the inhibitory population and the second excitatory population are active (corresponding to $p_{0,1,1}$, which is the stable fixed point in region 6). This result is confirmed by GLV equations (Figure~\ref{fig5}E). For an example that corresponds to region 4, we chose $a=0.98$ and $b=0.92$. In this case, the initial conditions for the membrane potentials of the first excitatory population (which will be high according to the mathematical analysis), were chosen randomly in the range of $0$ and $15\,\mathrm{mV}$. The membrane potentials of the inhibitory and the second excitatory populations were initially set to a random number between $0$ and $20\,\mathrm{mV}$. As indicated in Figure~\ref{fig7} F, and as predicted by the GLV equations in Figure~\ref{fig5}F, the activity transient corresponding to $p_{0,1,1}$ will eventually settle in $p_{1,0,1}$, meaning that the steady state of the network would have a highly active inhibitory and first excitatory subnetwork.

\begin{figure}[tbph]
        \centering
        \includegraphics[width=0.85\textwidth]{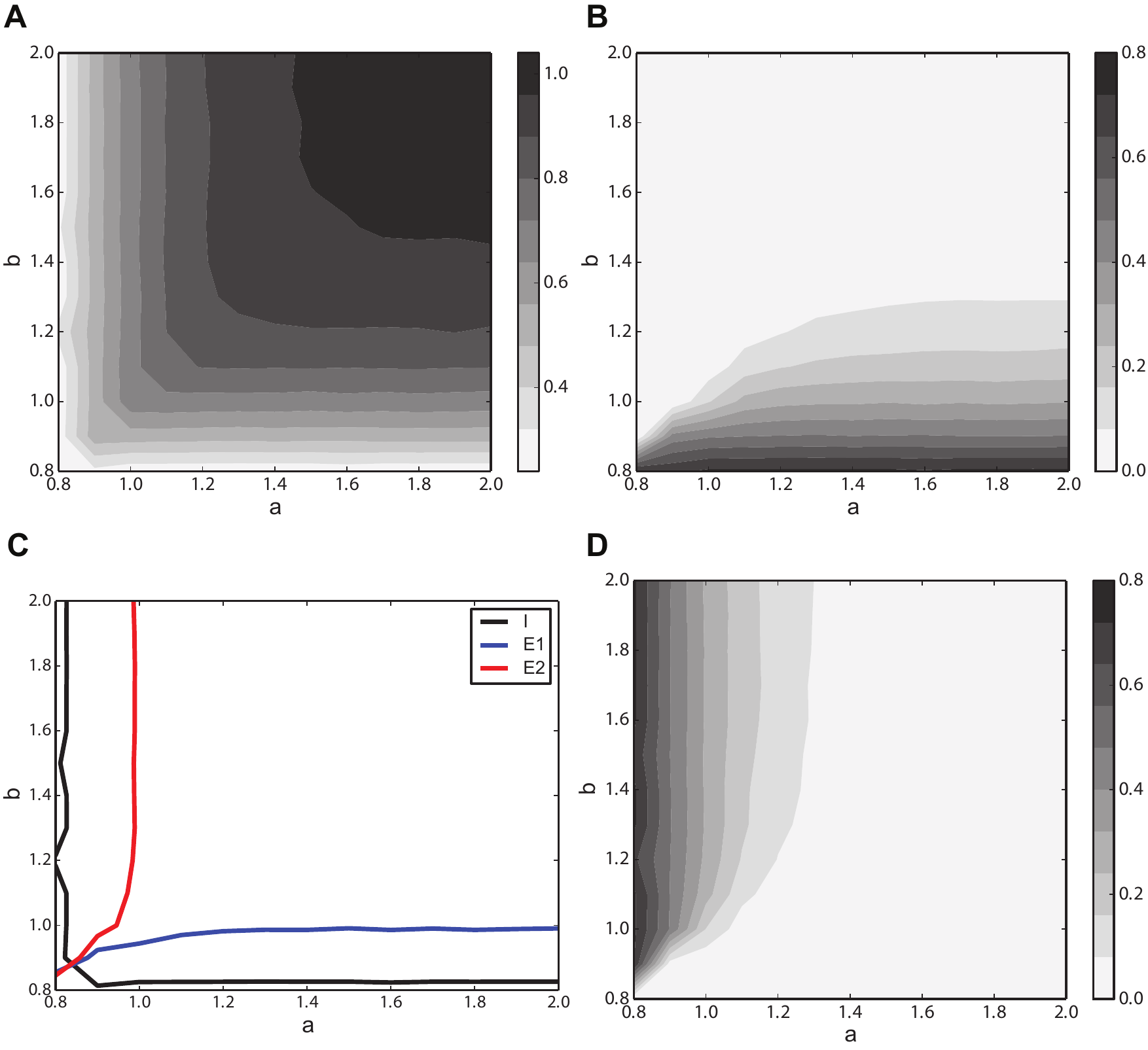}
        \caption{{\bf Bifurcation diagram of the collective activity of the network in the EEI scenario, extracted from numerical simulations.} In A, B, D, the relative ratios of the average neuronal firing rate for the inhibitory, first excitatory and second excitatory subnetwork are plotted, respectively. C: Contour lines of the ratio at a level of $0.3$ for all subnetworks.} 
\label{fig8}
\end{figure}

For $a$ and $b$ between $0.8$ and $2.0$, we performed network simulations for a duration of $4$ seconds, with a simulation time step of $0.1\,\mathrm{ms}$. After discarding the network transients to the steady state (the first $100\,\mathrm{ms}$), we plotted the ratio of the average neural firing rate in each population to the sum of the firing rates of individual neurons in each population (Figure~\ref{fig8} A-C). This plot provides sufficient information about the relative activities of the sub-populations. For values of $a$ and $b$ close to $2.0$, the average firing rate of an inhibitory neuron is large compared to the firing rates of the excitatory units. In these plots, values larger than $1/3$ indicate that the corresponding neuronal firing rate is bigger than the firing rates of other neurons in the other subnetworks. The bifurcation diagram of the network, obtained from network simulations, can intuitively be inferred from the contour lines of the relative firing rates, at the value of $1/3$. As presented in Figure~\ref{fig8} D, these contour lines, which can be interpreted as bifurcation lines (because they correspond to a different collective behavior), have a similar shape as those of Figure~\ref{fig3}. The reason is that in the specified parameter space, mostly transcritical bifurcations occur. This results in the change of stability of fixed points which manifests itself in the firing rate of the populations. In other words, when a change in firing rates (subnetwork state) occurs, it corresponds to a change of stability of the corresponding fixed point through a bifurcation. Compared to Figure~\ref{fig3}, it is clear that in the fluctuation driven regime, GLV can represent system dynamics with high fidelity. 
In the region where both $a$ and $b$ are less than $1$, the collective behavior of the network depends on the initial conditions. This is exactly the region where $p_{0,0,1}$ has two positive eigenvalues in the GLV model, which allows the trajectory to converge into any of the stable fixed points (where $a$ and $b$ are close to each other, corresponding to region 5 in Figure~\ref{fig3}), or into the unique stable fixed point (corresponding to region 4 or 6). 

\subsection*{III scenario}

\begin{figure}[tbph]
\centering
\includegraphics[width=0.6\textwidth]{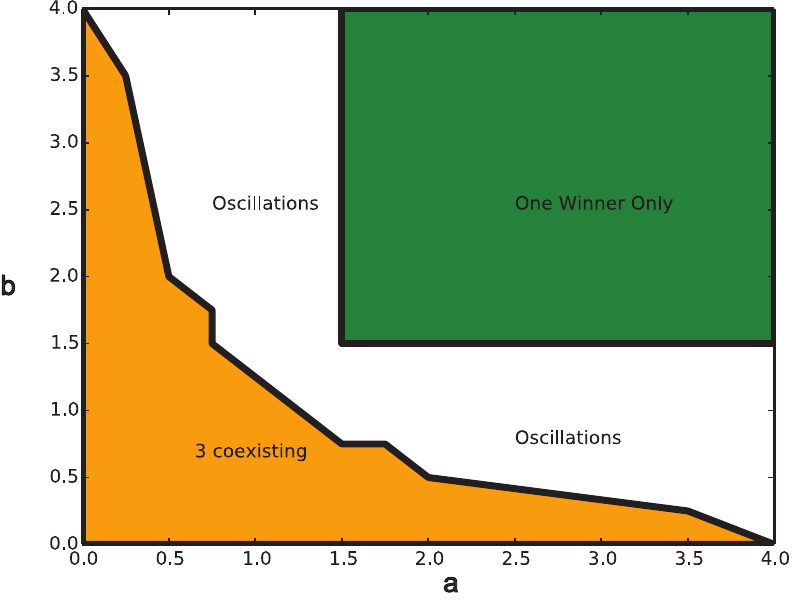}
\caption{{\bf Bifurcation diagram for the network dynamics in the III scenario.} Bifurcation diagram obtained from simulation results of a spiking network illustrates three different collective dynamics. Depending on the parameter combinations, either all subnetworks are active simultaneously (orange region), or only one population is active and others have a zero firing rate (green region), or the firing rates of the populations follow an oscillatory pattern (white region).}
\label{fig9}
\end{figure}

To demonstrate the power of GLV equations in representing spiking network dynamics, we also considered a purely inhibitory network that represents a May-Leonard competitive system \cite{mayleonard75}. We chose a synaptic coupling equal to $J=-0.012\,\mathrm{mV}$, but the network state is similar for larger values of $J$. The connection probability was $0.1$ throughout, and we considered networks with identical in-degrees and identical out-degrees for all neurons in order to exclude any structural bias for the network dynamics. The bifurcation diagram obtained by May and Leonard \cite{mayleonard75} through theoretical analysis has the following properties: (1) For $a+b<2$ the three inhibitory populations are active with equal rates. (2) For $a>1$ and $b>1$, only one population is active. (3) For any other parameter combination, the network shows oscillations in the firing rates of the three populations. The period of oscillations increases as a function of time; however, for $a+b=2$, a stable limit cycle solution exists. For details of the mathematical analysis, see \cite{mayleonard75}.
Figure~\ref{fig9} is obtained from spiking network simulations. There is a very good match between the two diagrams. For small values of $a$ and $b$, in the orange region of the diagram, all three neuronal populations are active, with non-zero firing rates. For values of $a$ and $b$ larger than $1.5$, only one population has a non-zero firing rate. In this diagram obtained from network simulations, there is a gap between the bifurcating regions (green and orange) that should disappear at $a,b=1$ (similar to \cite{mayleonard75}). Our simulations indicate that with increasing the network sizes, these small discrepancies vanish. For the parameter combinations in the white region of the figure, which exclude any of the the aforementioned behavior, an oscillatory dynamics was observed. 

\begin{figure}[tbph]
\centering
\includegraphics[width=1.0\textwidth]{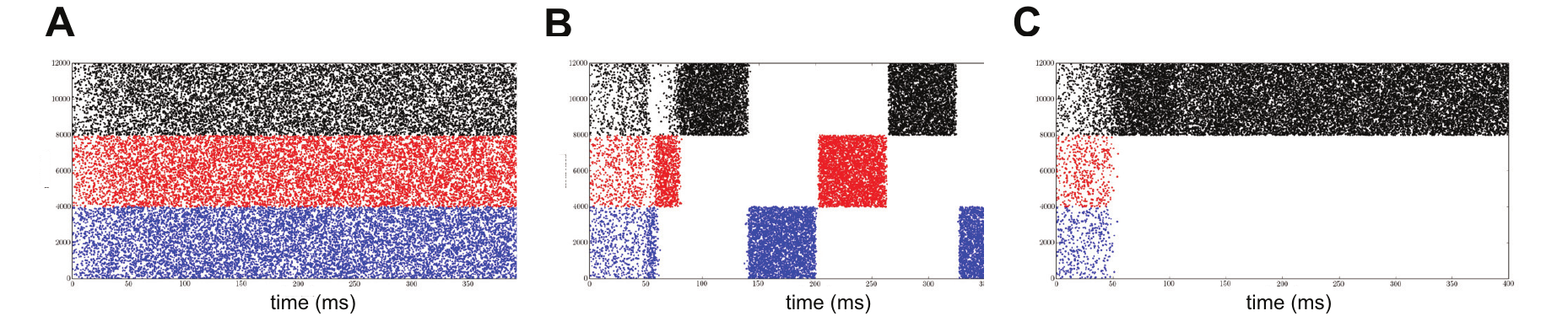}
\caption{{\bf Three different samples of spiking network simulations corresponding to the three different dynamical behaviors of the III network.} A: $a=0.75$, $b=0.75$, where all the populations are active. B: $a=2.25$, $b=1.25$, where the subnetworks oscillate with a phase difference of $2 \pi /3$. C: $a=2.0$, $b=2.0$, where only one inhibitory population is active, persistently dominating the other two populations.}
\label{fig10}
\end{figure}

There are three regions in the bifurcation diagram with different dynamics that are confirmed by simulation results in Figure~\ref{fig10}, which illustrates one example for each possible network dynamics. Note that in panel C of this figure, depending on the initial condition, the system has only one active inhibitory population, and all other populations remain silent.

In simulations for the parameter region where oscillations occur, we made the interesting observation that the period of the high activity of the three inhibitory populations increases as a function of time, following the initial transients (Figure~\ref{fig10}). However, the period does not appear to grow without bounds. We hypothesize that this phenomenon is due to finite-size fluctuations in the spiking network dynamics, which is not captured by the GLV. In other words, in simulations the population firing rate exhibits excursions that randomly deviate from the deterministic GLV solutions. This keeps the simulated trajectories to get very close to the heteroclinic cycle, avoiding the critical slowing down that would otherwise result. Therefore, the period of the oscillations cannot increase beyond a limit that is related to the amplitude of fluctuations of the network activity.    

\section*{Discussion}
We studied dynamical interactions between subnetworks of different types of neurons, within a large network. For two example networks, we focused on the role of the strength of couplings between subnetworks for global network dynamics. Such a study can shed new light, for example, on the interaction dynamics between different brain nuclei in the basal ganglia, or between ``columns'' or cell assemblies in a certain region of neocortex. Applications of such mathematical modeling would, among other things, help to predict aberrant network dynamics that underlies certain brain disorders, as described in Parkinson's Disease or certain types of epilepsy. 
 
In our study, the dynamics of networks comprising interacting subnetworks of excitatory and inhibitory neurons were compared with the dynamics of ecosystems comprised of prey and predator species, corresponding to the excitatory and inhibitory subnetworks, respectively.
Specifically, we considered Generalized Lotka-Volterra (GLV) systems and numerical simulations of spiking networks composed of three subnetworks, with leaky integrate-and-fire (LIF) neurons as dynamical nodes. In both cases, coupling strength was conceived as a bifurcation parameter. Bifurcation diagrams extracted from the GLV systems and from the numerical simulations were strikingly similar in the two examples we studied. This indicates that GLV represents a meaningful model of competing populations of spiking neurons, and there is a qualitative equivalence between the mathematical equations and the behavior of the simulated network. In the model considered here, it was possible to predict convergence towards the correct fixed points (as validated by the simulation results), as well as oscillatory dynamics in the purely inhibitory network, for the correct parameter regime. 

Whether GLV could represent a model for any network of arbitrary number of subnetworks is a question that needs further investigation. GLV equations, however, could be interpreted as a special case of Wilson-Cowan equations, wherein the population response function ($S(x)$ in their notation \cite{Wilson1972}), which is typically assumed to be a sigmoid function, is a linear function of the overall input (the overall excitation level that is fed into the network). Approximating this response function with a linear function may be valid only if the network is operating in a low firing rate asynchronous irregular regime.

There are, however, subtle differences between GLV equations that describe ecosystems, and the GLV we obtained to approximate the spiking network dynamics. In a two-prey-one-predator system, the coefficient of a prey population variable which affects the population growth of its own or any other prey, is usually taken negative \cite{Takeuchi1983}. This usually represents the competition between preys for limited resources. In the EEI network that we considered in this study, there is no direct competition between excitatory neurons for resources, which in this case is represented by the external input. Therefore, the influence of an excitatory population on its own activity or on the other excitatory population's activity is positive. As it turns out, for negative self-couplings of excitatory populations, and negative couplings between the two excitatory population, there exists a region in the parameter space where $p_{1,1,1}$ is a global attractor. This would result in a stable network state, where all subnetworks are active. 

For both sample configurations (EEI and IIE) considered in our paper, in the bifurcation diagram obtained from network simulations, there is a gap between the bifurcating regions that should disappear at $a,b=1$. Our simulations indicate that with increasing the network sizes, these small discrepancies vanish. Essentially, the larger the population size, the more precise the GLV system becomes as a low-dimensional description. Although relative population sizes were taken into account in our model, finite size effects and the exact scaling relations between the number of neurons in each subnetwork and the coefficients in the GLV equation are beyond the scope of this paper.
 
To summarize, Generalized Lotka-Volterra equations represent an interesting family of systems that can represent a wide dynamical repertoire, such as oscillations, sequential activities, and chaos. Therefore, they can be regarded as a good candidate to model dynamic interactions in neuronal networks on the population level.   
In this study, we did not aim at identifying the correct time scale of the dynamics from the network parameters, as its dependence is subtle. However, on a qualitative level, we could show that different strengths of couplings between subnetworks can lead to different particular behaviors, and we were able to validate these scenarios by neural network simulations. In fact, in biological systems like neuronal networks, the issue of time scales is under debate \cite{Marom2010, Duarte2017}. As our model is an abstract low-dimensional and simplified representation of a complex biological reality, we cannot claim to represent each and every phenomenon in our model. 

The bifurcation analysis in this paper paves the way for deterministic analysis of coupled networks in higher dimensions. It also could be used to study stochastic nonlinear dynamics of such systems, in different regimes of the network. However, in such cases, noise amplitudes can also play role as a bifurcation parameter.

GLV equations have been suggested as a framework to generate metastable systems with saddle points \cite{Afraimovich2008,Bick2010} that can exhibit winnerless competition dynamics. This framework allows for the emergence of robust transient dynamics \cite{Rabin}, which was hypothesized to underlie sensory encoding, for example in the olfactory system \cite{Rabinovich2008}. In fact, there is evidence that different odor stimuli trigger different transient trajectories and succession of states in a high dimensional neuronal response \cite{mazor2005,Laurent,Freeman2000}. Moreover, Generalized Lotka-Volterra equations have been implicated as models of various cognitive processes, such as decision-making, and sequential working memory \cite{Rabinovich2014a}. As a model for controlling the desynchronized phase of the sleep cycle, Lotka-Volterra equations have been suggested to replicate the dynamics of excitatory and inhibitory populations \cite{McCarley1975}. Also, to model the perception of color, such equations were applied to represent the dynamics of competing cortical neurons, which have wave-length dependent activity for redness, greenness, and short wave-length redness, to implement a winner-take-all representation of the phenomenon \cite{Billock2001}.
Justifying these equations for the collective behavior of spiking networks is an important step towards approaching a mathematical model for brain dynamics, bridging the gap between the different scales of analysis from small neuronal populations to global brain dynamics with a direct link to higher cognitive processes. This eventually will bring us closer to elucidating some fundamental principles of the brain's complex operations.

\section*{Acknowledgements}
Funding by the German Ministry of Education and Research (BMBF, grant BFNT 01GQ0830) and the German Research Foundation (DFG, grant EXC 1086) is gratefully acknowledged. The article processing charge was covered by the open access publication fund of the University of Freiburg.

\bibliographystyle{unsrt}
\bibliography{./MyCollection}

\begin{thebibliography}{10}

\bibitem{Mountcastle1997}
Vernon~B. Mountcastle.
\newblock {The columnar organization of the neocortex}.
\newblock {\em Brain}, 120(4):701--722, 1997.

\bibitem{Hebb1949}
D.~O. Hebb.
\newblock {\em {The Organization of Behavior; A Neuropsychological Theory}}.
\newblock 1949.

\bibitem{Harris2005}
Kenneth~D Harris.
\newblock {Neural signatures of cell assembly organization.}
\newblock {\em Nature reviews. Neuroscience}, 6(5):399--407, 2005.

\bibitem{Lansner2009}
Anders Lansner.
\newblock {Associative memory models: from the cell-assembly theory to
  biophysically detailed cortex simulations}.
\newblock {\em Trends in Neurosciences}, 32(3):178--186, 2009.

\bibitem{Abeles1991}
Moshe Abeles.
\newblock {\em {Corticonics : neural circuits of the cerebral cortex}}.
\newblock Cambridge University Press, 1991.

\bibitem{Abeles1982}
Moshe. Abeles.
\newblock {\em {Local Cortical Circuits : an Electrophysiological Study}}.
\newblock Springer Berlin Heidelberg, 1982.

\bibitem{kumar}
Ashok Litwin-Kumar and Brent Doiron.
\newblock {Formation and maintenance of neuronal assemblies through synaptic
  plasticity}.
\newblock {\em Nature Communications}, 5(May):5319, 2014.

\bibitem{Tetzlaff2015}
Christian Tetzlaff, Sakyasingha Dasgupta, Tomas Kulvicius, and Florentin
  W{\"{o}}rg{\"{o}}tter.
\newblock {The Use of Hebbian Cell Assemblies for Nonlinear Computation}.
\newblock {\em Scientific Reports}, 5:1--14, 2015.

\bibitem{Gallinaro2018}
J{\'{u}}lia~V. Gallinaro and Stefan Rotter.
\newblock {Associative properties of structural plasticity based on firing rate
  homeostasis in recurrent neuronal networks}.
\newblock {\em Scientific Reports}, 8(1):1--13, 2018.

\bibitem{Wilson1972}
H.~R. Wilson and J.~D. Cowan.
\newblock {Excitatory and inhibitory interactions in localized populations of
  model neurons.}
\newblock {\em Biophysical journal}, 12(1):1--24, jan 1972.

\bibitem{Stern1998}
E~.a Stern, D.~Jaeger, and C.~J. Wilson.
\newblock {Membrane potential synchrony of simultaneously recorded striatal
  spiny neurons in vivo.}
\newblock {\em Nature}, 394(6692):475--478, 1998.

\bibitem{Kumar2011}
Arvind Kumar, Stefano Cardanobile, Stefan Rotter, and Ad~Aertsen.
\newblock {The role of inhibition in generating and controlling Parkinson's
  disease oscillations in the Basal Ganglia.}
\newblock {\em Frontiers in systems neuroscience}, 5(October):86, 2011.

\bibitem{Hammond2007}
Constance Hammond, Hagai Bergman, and Peter Brown.
\newblock {Pathological synchronization in Parkinson's disease: networks,
  models and treatments}.
\newblock {\em Trends in Neurosciences}, 30(7):357--364, 2007.

\bibitem{Angulo-Garcia2016}
David Angulo-Garcia, Joshua~D Berke, and Alessandro Torcini.
\newblock {Cell assembly dynamics of sparsely-connected inhibitory networks: A
  simple model for the collective activity of striatal projection neurons}.
\newblock {\em PLoS Computational Biology}, 12(2):1--29, 2016.

\bibitem{Ponzi2010}
Adam Ponzi and Jeff Wickens.
\newblock {Sequentially switching cell assemblies in random inhibitory networks
  of spiking neurons in the striatum.}
\newblock {\em The Journal of neuroscience}, 30(17):5894--911, apr 2010.

\bibitem{VanVreeswijk1996a}
C.~van Vreeswijk and H.~Sompolinsky.
\newblock {Chaos in neuronal networks with balanced excitatory and inhibitory
  activity.}
\newblock {\em Science (New York, N.Y.)}, 274(5293):1724--1726, 1996.

\bibitem{Brunel2000}
Nicolas Brunel.
\newblock {Dynamics of sparsely connected networks of excitatory and inhibitory
  spiking neurons}.
\newblock {\em Journal of computational neuroscience}, 8:183--208, 2000.

\bibitem{Gerstner1995}
W.~Gerstner.
\newblock {Time structure of the activity in neural network models}.
\newblock {\em Physical Review E}, 51(1):738--758, 1995.

\bibitem{Schwalger2017}
Tilo Schwalger, Moritz Deger, and Wulfram Gerstner.
\newblock {\em {Towards a theory of cortical columns: From spiking neurons to
  interacting neural populations of finite size}}, volume~13.
\newblock 2017.

\bibitem{Volterra1926}
Vito Volterra.
\newblock {Variazioni e fluttuazioni del numero d'individui in specie animali
  conviventi}.
\newblock {\em Memorie della R. Accademia dei Lincei}, 6(2):31--113, 1926.

\bibitem{lotka}
Alfred~J. Lotka.
\newblock {Contribution to the theory of periodic reactions}.
\newblock {\em J. Phys. Chem}, 14:271--274, 1910.

\bibitem{Fukai1997a}
Tomoki Fukai.
\newblock {A Simple neural network exhibiting selective activation of neuronal
  ensembles: FromWinner-Take-All to Winners-Share-All}.
\newblock {\em Neural Computation}, 97:77--97, 1997.

\bibitem{Varona2002}
Pablo Varona, Mikhail~I. Rabinovich, Allen~I. Selverston, and Yuri~I.
  Arshavsky.
\newblock {Winnerless competition between sensory neurons generates chaos: A
  possible mechanism for molluscan hunting behavior}.
\newblock {\em Chaos}, 12(3):672--677, 2002.

\bibitem{Afraimovich2004}
V.~S. Afraimovich, V.~P. Zhigulin, and M.~I. Rabinovich.
\newblock {On the origin of reproducible sequential activity in neural
  circuits}.
\newblock {\em Chaos}, 14(4):1123--1129, 2004.

\bibitem{Cardanobile2010}
Stefano Cardanobile and Stefan Rotter.
\newblock {Multiplicatively interacting point processes and applications to
  neural modeling.}
\newblock {\em Journal of computational neuroscience}, 28(2):267--84, apr 2010.

\bibitem{Cardanobile2011}
Stefano Cardanobile and Stefan Rotter.
\newblock {Emergent properties of interacting populations of spiking neurons.}
\newblock {\em Frontiers in computational neuroscience}, 5(December):59, jan
  2011.

\bibitem{Grossberg}
Michael~A Cohen and Stephen Grossberg.
\newblock {Absolute stability of global pattern formation and parallel memory
  storage by competitive neural networks}.
\newblock {\em IEEE Transactions on Systems, Man, and Cybernetics}, SMC-13(5),
  1983.

\bibitem{Li2014}
Jun Li, Jian Yang, Xiaotong Yuan, and Zhaohua Hu.
\newblock {Continuous attractors of higher-order recurrent neural networks with
  infinite neurons}.
\newblock {\em Neurocomputing}, 131(10):388--396, 2014.

\bibitem{Hofbauer1994}
J~Hofbauer.
\newblock {Multiple limit cycles for three dimensional Lotka- Volterra
  equations}.
\newblock {\em Appl. Math. Lett.}, 7(6):65--70, 1994.

\bibitem{gilpin}
Michael~E. Gilpin.
\newblock {Spiral chaos in a predator- prey model}.
\newblock {\em The American Naturalist}, 113(2):306--308, 1977.

\bibitem{Rabinovich2008}
Misha Rabinovich, Ramon Huerta, and Gilles Laurent.
\newblock {Transient dynamics for neural processing.}
\newblock {\em Science}, 321(5885):48--50, 2008.

\bibitem{rabin2001}
M.~Rabinovich, A.~Volkovskii, P.~Lecanda, R.~Huerta, H.~D. Abarbanel, and
  G.~Laurent.
\newblock {Dynamical encoding by networks of competing neuron groups:
  winnerless competition.}
\newblock {\em Physical review letters}, 87:068102, 2001.

\bibitem{mazor2005}
Ofer Mazor and Gilles Laurent.
\newblock {Transient dynamics versus fixed points in odor representations by
  locust antennal lobe projection neurons}.
\newblock {\em Neuron}, 48(November 23):661--673, 2005.

\bibitem{Perin2011}
Rodrigo Perin, Thomas~K. Berger, and Henry Markram.
\newblock {A synaptic organizing principle for cortical neuronal groups}.
\newblock {\em Proceedings of the National Academy of Sciences of the United
  States of America}, 108(13):5419--24, mar 2011.

\bibitem{mayleonard75}
Robert~M. May and Warren~J. Leonard.
\newblock {Nonlinear aspects of competition between three species}.
\newblock {\em SIAM Journal on Applied Mathematics}, 29(2):243--253, 1975.

\bibitem{Lagzi2014}
Fereshteh Lagzi and Stefan Rotter.
\newblock {A Markov model for the temporal dynamics of balanced random networks
  of finite size}.
\newblock {\em Frontiers in Computational Neuroscience}, 8(December):1--23, dec
  2014.

\bibitem{Newman2001}
M.~E. Newman, S.~H. Strogatz, and D.~J. Watts.
\newblock {Random graphs with arbitrary degree distributions and their
  applications.}
\newblock {\em Physical review. E, Statistical, nonlinear, and soft matter
  physics}, 64(2 Pt 2):026118, 2001.

\bibitem{Hofstad2014}
Remco van~der Hofstad.
\newblock {\em {Random graphs and vomplex networks}}, volume~1.
\newblock 2014.

\bibitem{Dhooge2003}
A.~Dhooge, W.~Govaerts, and Yuri.~A. Kuznetsov.
\newblock {MATCONT: A MATLAB package for numerical bifurcation analysis of
  ODEs}.
\newblock {\em ACM Transactions on Mathematical Software}, 29(2):141--164,
  2003.

\bibitem{strogatz2000}
Steven~H Strogatz.
\newblock {\em {Nonlinear Dynamics and Chaos: With Applications to Physics,
  Biology, Chemistry and Engineering}}.
\newblock Westview Press, 2000.

\bibitem{Gewaltig2007}
Marc-Oliver Gewaltig and Markus Diesmann.
\newblock {NEST (NEural Simulation Tool)}.
\newblock {\em Scholarpedia}, 2(4):1430, apr 2007.

\bibitem{Kuznetsov2004}
Yuri.~A. Kuznetsov.
\newblock {\em {Elements of applied bifurcation theory}}.
\newblock Springer, 3rd edition, 2004.

\bibitem{lagzi2015}
Fereshteh Lagzi and Stefan Rotter.
\newblock {Dynamics of competition between sub-networks of spiking neuronal
  networks in the balanced state}.
\newblock {\em submitted to PLoS One}, pages 1--29, 2015.

\bibitem{Takeuchi1983}
Yasuhiro Takeuchi and Norihiko Adachi.
\newblock {Existence and bifurcation of stable equilibrium in two-prey,
  one-predator communities}.
\newblock {\em Bulletin of Mathematical Biology}, 45(6):877--900, 1983.

\bibitem{Marom2010}
Shimon Marom.
\newblock {Neural timescales or lack thereof}.
\newblock {\em Progress in Neurobiology}, 90(1):16--28, 2010.

\bibitem{Duarte2017}
Renato Duarte, Alexander Seeholzer, Karl Zilles, and Abigail Morrison.
\newblock {Synaptic patterning and the timescales of cortical dynamics}.
\newblock {\em Current Opinion in Neurobiology}, 43(April):156--165, 2017.

\bibitem{Afraimovich2008}
Valentin Afraimovich, Irma Tristan, Ramon Huerta, and Mikhail~I. Rabinovich.
\newblock {Winnerless competition principle and prediction of the transient
  dynamics in a Lotka-Volterra model}.
\newblock {\em Chaos}, 18(2008), 2008.

\bibitem{Bick2010}
Christian Bick and Mi~Rabinovich.
\newblock {On the occurrence of stable heteroclinic channels in
  Lotka–Volterra models}.
\newblock {\em Dynamical Systems}, 25(1):37--41, 2010.

\bibitem{Rabin}
Mikhail~I Rabinovich and Pablo Varona.
\newblock {Robust transient dynamics and brain functions}.
\newblock {\em Frontiers in Computational Neuroscience}, 5, 2011.

\bibitem{Laurent}
Gilles Laurent, Mark Stopfer, Rainer~W Friedrich, Misha~I Rabinovich, Alexander
  Volkovskii, and Henry D~I Abarbanel.
\newblock {Odor encoding as an active, dynamical process: experiments,
  computation, and theory}.
\newblock {\em Annu. Rev. Neurosci}, 24:263(97), 2001.

\bibitem{Freeman2000}
W.~J. Freeman and J.~M. Barrie.
\newblock {Analysis of spatial patterns of phase in neocortical gamma EEGs in
  rabbit.}
\newblock {\em Journal of neurophysiology}, 84(3):1266--1278, 2000.

\bibitem{Rabinovich2014a}
Mikhail~I. Rabinovich, Yury Sokolov, and Robert Kozma.
\newblock {Robust sequential working memory recall in heterogeneous cognitive
  networks}.
\newblock {\em Frontiers in Systems Neuroscience}, 8(November):1--11, 2014.

\bibitem{McCarley1975}
Robert~W McCarley and J~Allan Hobson.
\newblock {Neuronal excitability modulation over the sleep cycle : A structural
  and mathematical model}.
\newblock {\em Science}, 189(4196):58--60, 1975.

\bibitem{Billock2001}
V~a Billock, G~a Gleason, and B~H Tsou.
\newblock {Perception of forbidden colors in retinally stabilized equiluminant
  images: an indication of softwired cortical color opponency?}
\newblock {\em Journal of the Optical Society of America. A, Optics, image
  science, and vision}, 18(10):2398--2403, 2001.

\end{thebibliography}

\end{document}